\documentclass[aps,prd,reprint,superscriptaddress]{revtex4-2}

\usepackage[utf8]{inputenc}
\usepackage[T1]{fontenc}
\usepackage{graphicx}
\usepackage{siunitx}
\usepackage{subcaption}
\usepackage[colorlinks=true,
            linkcolor=blue,
            citecolor=blue,
            urlcolor=blue]{hyperref}
\usepackage{url}
\usepackage{booktabs}
\usepackage{amsmath,amsfonts}
\usepackage{nicefrac}
\usepackage{microtype}
\usepackage{xcolor}
\usepackage{listings}
\usepackage{listings-ext}

\begin{document}

\makeatletter
\renewcommand\@makecaption[2]{%
  \small\rmfamily
  \begin{list}{}{%
      \setlength{\leftmargin}{0pt}%
      \setlength{\rightmargin}{0pt}%
    }%
    \item[]\textbf{#1:} #2%
  \end{list}%
}
\makeatother

\title{Automating High Energy Physics Data Analysis with LLM-Powered Agents}






\author{Eli Gendreau-Distler}
\email{egendreaudistler@berkeley.edu}
\affiliation{Department of Physics, University of California, Berkeley, Berkeley, CA 94720, USA}
\affiliation{Physics Division, Lawrence Berkeley National Laboratory, Berkeley, CA 94720, USA}

\author{Joshua Ho}
\email{ho22joshua@berkeley.edu}
\affiliation{Department of Physics, University of California, Berkeley, Berkeley, CA 94720, USA}
\affiliation{Physics Division, Lawrence Berkeley National Laboratory, Berkeley, CA 94720, USA}

\author{Dongwon Kim}
\email{dwkim@berkeley.edu}
\affiliation{Department of Physics, University of California, Berkeley, Berkeley, CA 94720, USA}
\affiliation{Physics Division, Lawrence Berkeley National Laboratory, Berkeley, CA 94720, USA}

\author{Luc Tomas Le Pottier}
\email{luclepot@berkeley.edu}
\affiliation{Department of Physics, University of California, Berkeley, Berkeley, CA 94720, USA}
\affiliation{Physics Division, Lawrence Berkeley National Laboratory, Berkeley, CA 94720, USA}

\author{Haichen Wang}
\email{haichenwang@berkeley.edu}
\affiliation{Department of Physics, University of California, Berkeley, Berkeley, CA 94720, USA}
\affiliation{Physics Division, Lawrence Berkeley National Laboratory, Berkeley, CA 94720, USA}

\author{Chengxi Yang}
\email{cxyang@berkeley.edu}
\affiliation{Department of Physics, University of California, Berkeley, Berkeley, CA 94720, USA}
\affiliation{Physics Division, Lawrence Berkeley National Laboratory, Berkeley, CA 94720, USA}

\begin{abstract}
We present a proof-of-principle study demonstrating the use of large language model (LLM) agents to automate a representative high energy physics (HEP) analysis. Using the Higgs boson diphoton cross-section measurement as a case study with ATLAS Open Data, we design a hybrid system that combines an LLM-based supervisor–coder agent with the \texttt{Snakemake} workflow manager. In this architecture, the workflow manager enforces reproducibility and determinism, while the agent autonomously generates, executes, and iteratively corrects analysis code in response to user instructions. We define quantitative evaluation metrics -- success rate, error distribution, costs per specific task, and average number of API call for the task -- to assess agent performance across multi-stage workflows. To characterize variability across architectures, we benchmark a representative selection of state-of-the-art LLMs, spanning the \textit{Gemini} and \textit{GPT-5} series, the \textit{Claude} family, and leading open-weight models.  While the workflow manager ensures deterministic execution of all analysis steps, the final outputs still show stochastic variation. Although we set the temperature to zero, other sampling parameters (e.g., top-p, top-k) remained at their defaults, and some reasoning-oriented models internally adjust these settings. Consequently, the models do not produce fully deterministic results. This study establishes the first LLM-agent-driven automated data-analysis framework in HEP, enabling systematic benchmarking of model capabilities, stability, and limitations in real-world scientific computing environments. The baseline code used in this work is available at \href{https://huggingface.co/HWresearch/LLM4HEP}{https://huggingface.co/HWresearch/LLM4HEP}. This work was accepted as a poster at the Machine Learning and the Physical Sciences (ML4PS) workshop of the 39th Conference on Neural Information Processing Systems (NeurIPS 2025). The initial submission was made on August 30, 2025.
\end{abstract}
\maketitle

\section{Introduction}
Large language models (LLMs) and agent-based systems are increasingly being explored in scientific computing, with applications appearing in areas such as genomics and software engineering~\citep{xiao2024cellagentllmdrivenmultiagentframework,wang2025empiricalresearchutilizingllmbased}. 
Despite the inherently structured nature of collider-related data analyses and the potential advantages of automated, reproducible workflows, the usage of LLMs in high energy physics (HEP) is still at an early stage. 
Existing efforts in HEP have focused mainly on event-level inference or on domain-adapted language models~\citep{richmond2025feyntunelargelanguagemodels,simons2024astrohepbertbidirectionallanguagemodel,wu2024scalingparticlecollisiondata}, leaving open questions about how LLMs might participate directly in end-to-end analysis procedures.

Related work in agentic methodologies, including the \textit{Agents of Discovery} study~\citep{Diefenbacher:2025zzn}, demonstrates growing interest in structuring collider physics tasks through modular agent behavior. While such efforts examine agent-based reasoning in HEP contexts, they address different objectives and operational settings. Our study is situated within a workflow-management environment and considers how agentic components may be incorporated into a directed, reproducible analysis pipeline.

We introduce a framework in which LLM-based agents are integrated into a \texttt{Snakemake}-managed workflow~\citep{molder2021snakemake}. Agent interventions are bounded to well-defined tasks such as code generation, event selection, and validation while the underlying directed acyclic graph (DAG) maintains determinism and provenance. This design enables a controlled evaluation of the practicality and reliability of agent-driven steps within a full collider-analysis setting.

\section{Background and Motivation}
Research connecting LLMs with HEP is rapidly developing, though the existing literature remains concentrated on a small number of application areas -- for instance, enhancements to event-level prediction tasks, improvements to simulation, or tools for navigating domain knowledge. A substantial body of work applies transformer-based architectures to established HEP tasks such as classification, regression, and generative modeling~\citep{Qu:2022mxj,Qiu:2022xvr,Huang:2024voo,Mikuni:2024qsr}. These studies demonstrate the effectiveness of modern model architectures within conventional workflows but are not designed to automate or restructure the broader analysis process.

LLMs have also been used to improve access to experiment-specific documentation and technical resources. Systems such as \texttt{Chatlas}~\citep{DalSanto:2935252}, \texttt{LLMTuner}~\citep{schnabel2025largelanguagemodelsnew}, and \texttt{Xiwu}~\citep{xiwu} demonstrate the usefulness of natural-language interfaces for accessing experiment-specific documentation and technical information. Their focus, however, is on retrieval and interpretation rather than direct interaction with analysis pipelines or workflow tools.

A further line of research examines the use of LLMs for code generation and the operation of scientific software. Examples include ChatGPT-assisted analysis scripting at the Electron–Ion Collider~\citep{eic} and the use of LLMs to configure and run FLUKA simulations~\citep{ndum2024autoflukalargelanguagemodel,fluka}. These demonstrations highlight that LLMs can interface with complex software environments, motivating exploration of more integrated, workflow-level applications.

Developments outside HEP provide additional context for an agent-based approach. Structured prompting has enabled LLMs to execute multi-step scientific procedures by decomposing tasks into modular operations~\citep{Pan2025LLMHF,reasoner,benchmark}. In parallel, lightweight and edge-deployable models - illustrated by the \texttt{TinyAgent} framework~\citep{tinyagent} and models such as TinyLlama-1.1B~\citep{tinyllama} and Wizard 2-7B~\citep{wizard} have shown that smaller models can produce reliable action plans while offering advantages in latency, privacy, and local deployability~\citep{hammer,hera,dot}.

Together, these developments indicate a timely opportunity to study how agentic components might complement existing HEP workflows, particularly within reproducible, directed execution environments. The present work is motivated by this setting and aims to evaluate, in a controlled manner, how LLM-based agents perform when embedded into a structured collider-analysis pipeline. 

\section{Description of a representative high energy physics analysis}\label{sec:description_HEP_analysis}
In this paper, we use a cross-section measurement of the Higgs boson decaying to two photons at the Large Hadron Collider as an example. We employ collision data and simulation samples from the 2020 ATLAS Open Data release~\citep{ATLAS2020SimSamples}. The data sample corresponds to $10\,\mathrm{fb}^{-1}$ of proton-proton collisions collected in 2016 at $\sqrt{s} = 13$~TeV. Simulated samples include higgs boson production via gluon-gluon fusion, vector boson fusion, associated production with a vector boson, and associated production with a top-quark pair. Our example analysis uses control samples derived from collision data and Higgs production simulation samples to design a categorized analysis with a machine-learning-based event classifier, similar to the one used by the CMS collaboration in their Higgs observation paper~\citep{CMS:2012qbp}. The objective of this analysis is to maximize the expected significance of the Higgs-to-diphoton signal. Since the purpose of this work is to demonstrate the LLM-based agentic implementation of a HEP analysis, we do not report the observed significance. The workflow of this analysis is representative of resonance searches, a common analysis signature in HEP. 

The technical implementation of the analysis workflow is factorized into five sequential steps, executed through the \texttt{Snakemake} workflow management system. Each step is designed to test a distinct type of reasoning or code-generation capability, and the evaluation of each step is performed independently, i.e., the success of a given step does not depend on the completion or correctness of the previous one. This design allows for consistent benchmarking across heterogeneous tasks while maintaining deterministic workflow execution.

\textbf{Step 1} (\textbf{ROOT file inspection}) generates summary text files describing the ROOT files and their internal structure, including trees and branches, to provide the agent with a human-readable overview of the available data.

\textbf{Step 2} (\textbf{Ntuple conversion}) produces a Python script that reads all particle observables specified in the user prompt from the \texttt{TTree} objects in ROOT files~\citep{Antcheva:2009zz} and converts them into \texttt{numpy} arrays for downstream analysis.

\textbf{Step 3} (\textbf{Preprocessing}) normalizes the signal and background arrays and applies standard selection criteria to prepare the datasets for machine-learning–based classification.

\textbf{Step 4} (\textbf{S-B separation}) applies TabPFN~\citep{Hollmann2025TabularFM}, a transformer-based foundation model for tabular data, to perform signal-background (S-B) separation. The workflow requires a script that calls a provided function to train and evaluate the TabPFN model using the appropriate datasets and hyperparameters.

\textbf{Step 5} (\textbf{Categorization}) performs statistical categorization of events by defining optimized boundaries on the TabPFN score to maximize the expected significance of the Higgs-to-diphoton signal. This step is implemented as an iterative function-calling procedure, where new category boundaries are added until the improvement in expected significance falls below 5\%.

Additional physics-motivated details for steps 3, 4, and 5 are provided in Appendix~\ref{appendix:physics_details}.

\section{Architecture}

We adopt a hybrid approach to automate the Higgs boson to diphoton data analysis. Given the relatively fixed workflow, we use \texttt{Snakemake} to orchestrate the sequential execution of analysis steps. A supervisor–coder agent is deployed to complete each step. This design balances the determinism of the analysis structure with the flexibility often required in HEP analyses. 

\subsection{Workflow management}
\texttt{Snakemake} is a Python-based workflow management system that enables researchers to define computational workflows through rule-based specifications, where each rule describes how to generate specific output files from input files using defined scripts or commands. \texttt{Snakemake} serves as the workflow orchestration engine that manages the complex dependencies and execution order of the HEP analysis pipeline.  In this package, \texttt{Snakemake} operates as the backbone that coordinates the five-stage analysis workflow for ATLAS diphoton data processing. This modular approach allows the complex physics analysis to be broken down into manageable, interdependent components that can be executed efficiently and reproducibly.

\subsection{Supervisor-coder agent}
\begin{figure*}[htbp]
  \centering
  \includegraphics[width=0.95\textwidth]{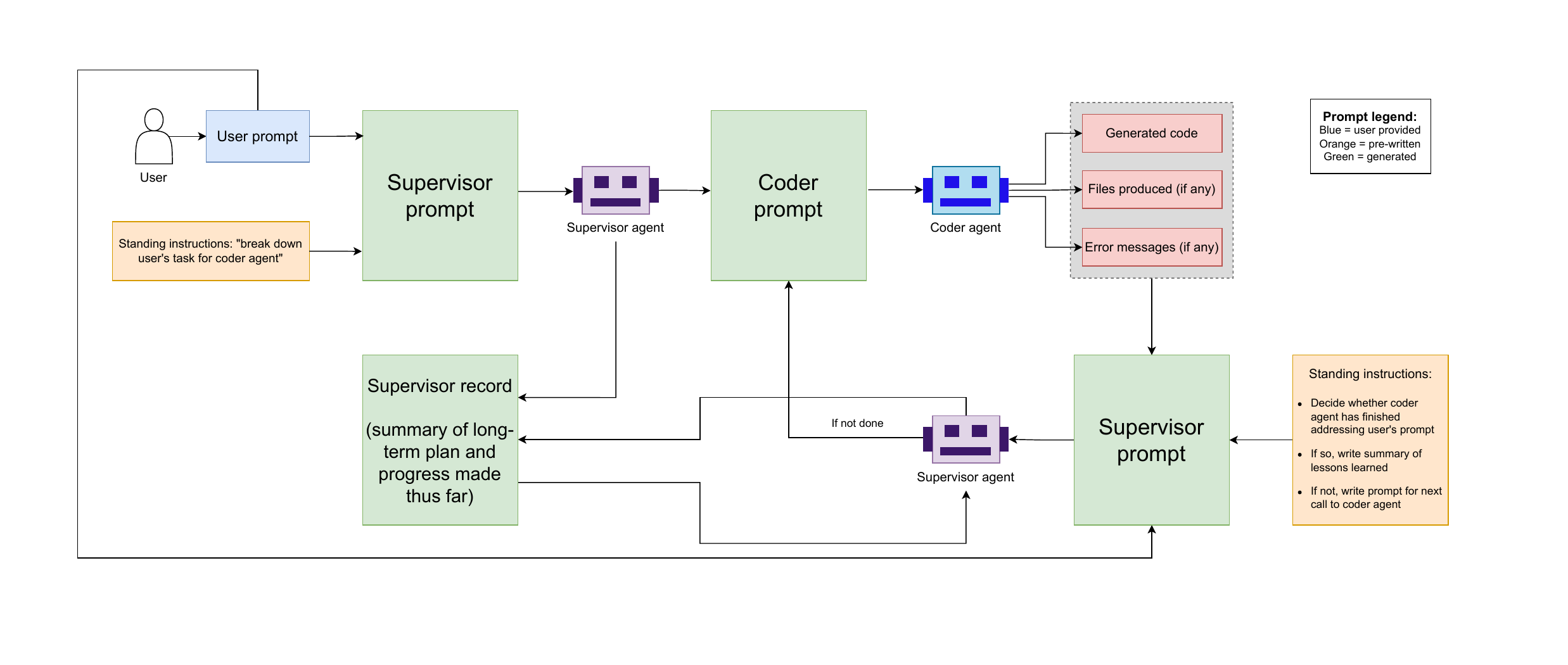}
  \caption{
  \textbf{Illustration of the internal workflow for the supervisor–coder agent}. The blue box denotes user-provided prompts, orange boxes denote pre-written standing instructions (system prompts), and green boxes denote dynamically generated prompts created during the workflow. The supervisor decomposes the user’s task into subtasks, maintains an evolving supervisor record, and issues structured prompts to the coder agent. The coder returns generated code, intermediate files, and error messages, which the supervisor evaluates to determine whether the task has been completed or whether further refinement is required. This iterative supervisor–coder loop continues until the coder produces a valid solution or the maximum number of allowed attempts is reached.
}
  \label{fig:supervisor-coder-agent}
\end{figure*}

We design a supervisor-coder agent to carry out each task, as illustrated in Fig.~\ref{fig:supervisor-coder-agent}. The supervisor and coder are implemented as API calls to a LLM, with distinct system prompts tailored to their respective roles. The supervisor receives instructions from the human user, formulates corresponding directives for the coder, and reviews the coder’s output. The coder, in turn, takes the supervisor’s instructions and generates code to implement the required action. The generated code is executed through an execution engine, which records the execution trace. The supervisor and coder roles are defined by their differing access to state, memory, and system instructions. In the reference configuration, both roles are implemented using the \texttt{gemini-pro-2.5}~\cite{gemini_pro} model; however, the same architecture is applied to a range of contemporary LLMs to evaluate model-dependent performance and variability.

Each agent interaction is executed through API calls to the LLM. Although we set the temperature to 0, other sampling parameters (e.g., top-p, top-k) remained at their default values, and some thinking-oriented models internally raise the effective temperature. As a result, the outputs exhibit minor stochastic variation even under identical inputs. Each call includes a user instruction, a system prompt, and auxiliary metadata for tracking errors and execution records.

For the initial user interaction with the supervisor, the input prompt includes a natural-language description of the task objectives, suggested implementation strategies, and a system prompt that constrains the behavior of the model output. The result of this interaction is an instruction passed to the coder, which in turn generates a Python script to be executed. If execution produces an error, the error message, the original supervisor instruction, the generated code, and a debugging system prompt are passed back to the supervisor in a follow-up API call. The supervisor then issues revised instructions to the coder to address the problem. This self-correction loop is repeated up to three times, after which the trial is marked unsuccessful and the process stops.

\section{Results}
\subsection{Statistics-enriched}
To establish a baseline, we conducted 219 experiments with the \texttt{gemini-pro-2.5} model, providing high statistical precision across all analysis stages. In this configuration, the workflow was organized into three composite steps: (1) \textbf{data preparation}, including \textit{ROOT file inspection}, \textit{ntuple conversion}, and \textit{preprocessing}; (2) \textbf{signal-background (S-B) separation}; and (3) \textbf{categorization} based on the expected Higgs-to-diphoton significance. Up to five self-correction iterations were permitted, as preliminary experiments showed that additional attempts rarely improved outcomes while incurring significantly higher computational cost. The corresponding success rates for these three steps were 58~$\pm$ 3\%, 88~$\pm$ 2\%, and 74~$\pm$ 3\%, respectively.

\begin{table}[htbp]
\centering
\resizebox{\linewidth}{!}{%
\begin{tabular}{l S[table-format=3.0] S[table-format=3.0] S[table-format=3.0]}
\hline
\textbf{Error type} & \textbf{Data-prep} & \textbf{S--B sep.} & \textbf{Categorization} \\
\hline\hline
All data weights = 0        & 41 & 0  & 0  \\
Dummy data created          & 15 & 4  & 4  \\
Function-calling error      & 1  & 3  & 26 \\
Incorrect branch name       & 9  & 0  & 0  \\
Intermediate file not found & 17 & 13 & 6  \\
Semantic error              & 6  & 5  & 17 \\
Other                       & 4  & 0  & 4  \\
\hline
\textbf{Total failures}     & \textbf{93} & \textbf{25} & \textbf{57} \\
\hline
\end{tabular}
} 

\vspace{6pt}
\caption{
Distribution of failure counts by error category across the three workflow stages, based on 219 trials per step. Each row lists the number of failures attributable to a given error type; the bottom row reports the total failures per stage.
}
\label{tab:error_types}
\end{table}
To analyze failure modes, we examined the model-generated log files for all unsuccessful trials and grouped recurrent issues into seven error categories. These categories correspond to the most frequently observed patterns across the logs and reflect dominant failure modes in agent-driven HEP workflows. As summarized in Table~\ref{tab:error_types}, the \textbf{data-preparation} (data-prep) stage was the most error-prone, with 93 failures out of 219 trials. Most issues arose from insufficient context for identifying objects within \textsc{ROOT} files. The most common error types - \textit{all data weights = 0} and \textit{intermediate file not found} - indicate persistent challenges in reasoning about file structures, dependencies, and runtime organization. Providing richer execution context, such as package lists, file hierarchies, and metadata, may help mitigate these problems. Despite these limitations, the overall completion rate ($\sim$57\%) demonstrates that LLMs can autonomously execute complex domain-specific programming tasks a nontrivial fraction of the time.

The subsequent \textbf{S–B separation} (S--B sep.) and \textbf{categorization} stages exhibited substantially fewer failures (25 and 57, respectively). In the S--B sep. step, the most common issue was \textit{intermediate file not found}, indicating that the agent occasionally lost track of file paths or dependencies across iterations. Once the workflow progressed to the \textbf{categorization} stage, failures were instead dominated by \textit{function-calling errors} and \textit{semantic errors}. This shift reflects the increased complexity of the categorization logic: the agent must iteratively construct valid function calls, interpret model outputs, and reason about numerical optimization criteria. These operations rely more heavily on precise syntax and conceptual understanding, making them more susceptible to calling and semantic mistakes despite the greater overall stability of later stages.

To assess agent efficiency, we measured the ratio of user-input tokens to the total tokens exchanged with the API. For the \texttt{gemini-pro-2.5} baseline, this ratio was (1.65~$\pm$0.15)$\times 10^{-3}$ for \textbf{data preparation}, (1.43$\pm$0.10)$\times 10^{-3}$ for \textbf{S–B separation}, and (0.93$\pm$~0.07)$\times 10^{-3}$ for \textbf{categorization}. A higher ratio indicates more efficient task execution, as fewer internally generated tokens are required to complete the same instruction set. Over 98\% of tokens originated from the model’s autonomous reasoning and self-correction rather than direct user input, suggesting that providing additional prompt detail would minimally affect overall token cost. This ratio therefore serves as a simple diagnostic of communication efficiency and reasoning compactness.

\subsection{LLM benchmarks}
\begin{figure*}[htbp]
  \centering

  \begin{subfigure}{0.7\textwidth}   
    \hspace{-3.0cm}
    \centering
    \includegraphics[width=\textwidth]{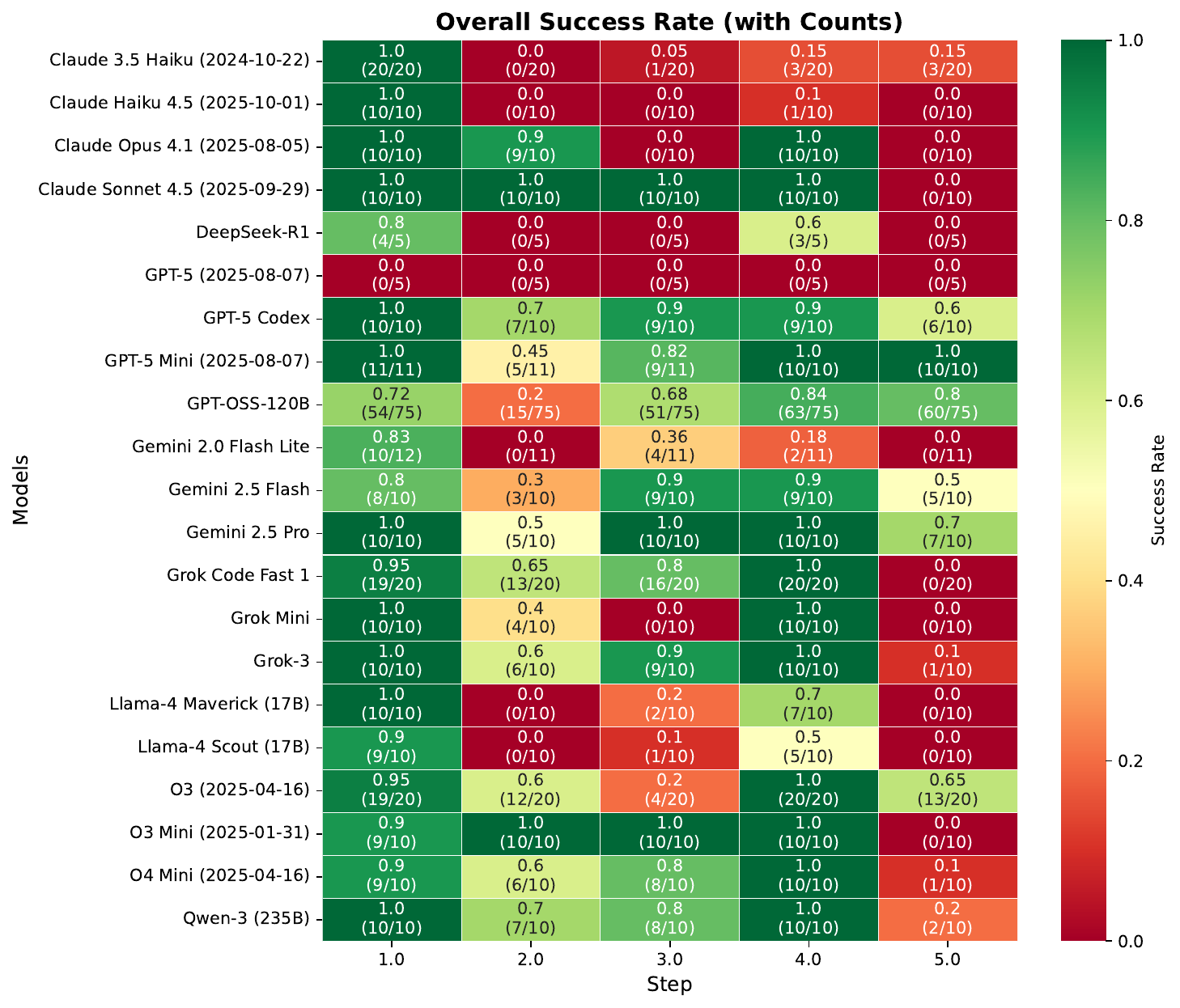}
    \caption{Success rate}
    \label{fig:success_heatmap}
  \end{subfigure}

  \vspace{1.2em}

  \begin{subfigure}{0.75\textwidth}   
    \centering
    \includegraphics[width=\textwidth]{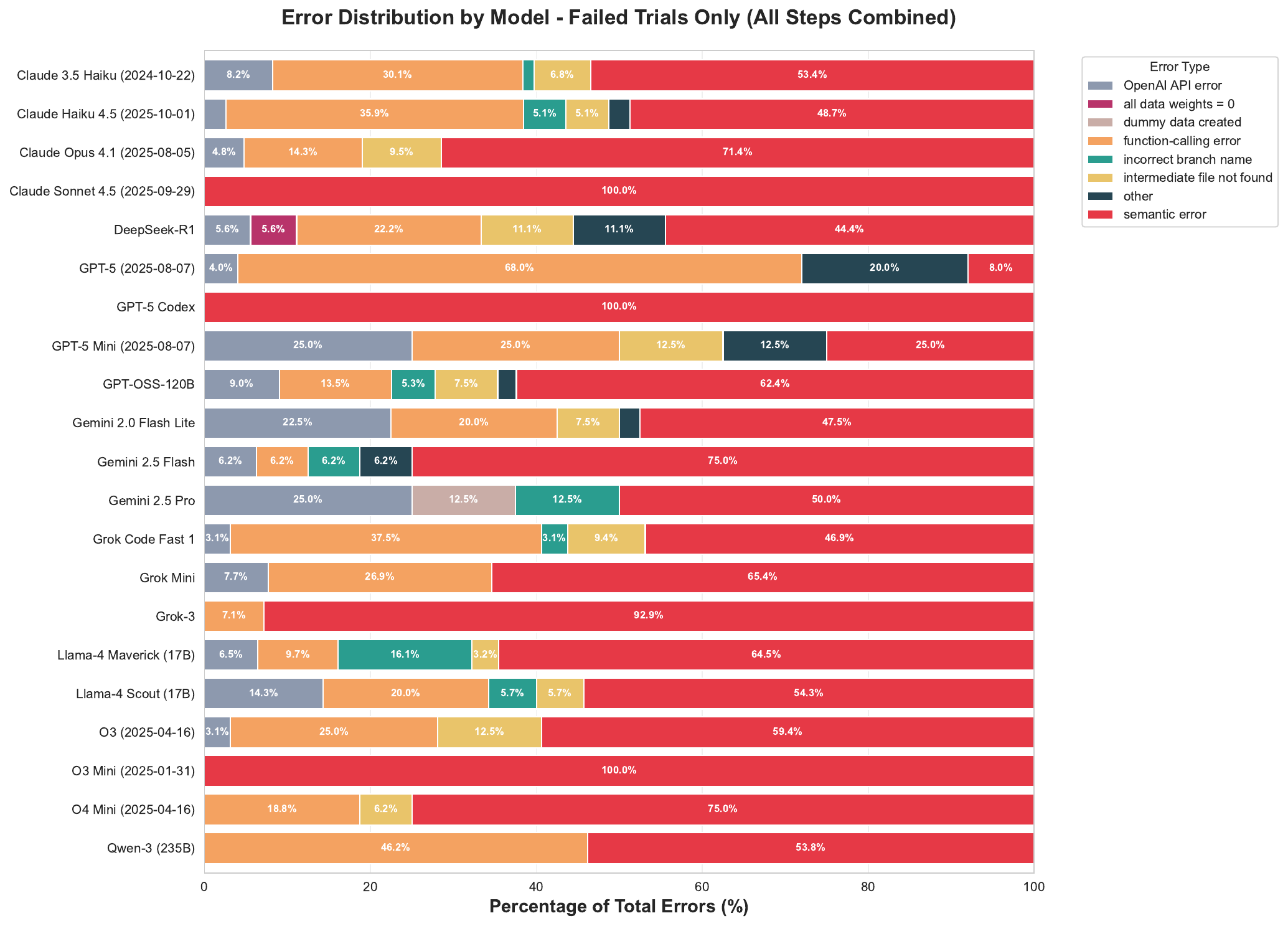}
    \caption{Error distribution}
    \label{fig:error_distribution}
  \end{subfigure}

  \caption{
    Cross-model comparison of LLM-agent performance.
    (a) Success fraction for each model-step pair.
    (b) Error distribution across all failed trials for each model.
  }
  \label{fig:model_comparison}
\end{figure*}





Following the initial benchmark with \texttt{gemini-pro-2.5}, we expanded the study to include additional models, such as \texttt{openai-gpt-5}~\citep{gpt5}, \texttt{claude-3.5}~\citep{anthropic2024claude35}, \texttt{qwen-3}~\citep{qwen3}, and the open-weight \texttt{gpt-oss-120b}~\citep{openai2025gptoss120bgptoss20bmodel}, evaluated under the same agentic workflow. Based on early observations, the prompt for the \textbf{data-preparation} stage was refined and subdivided into three subtasks: \textbf{ROOT file inspection}, \textbf{ntuple conversion}, and \textbf{preprocessing} (including signal and background region selection), as described in Section~\ref{sec:description_HEP_analysis}. Input file locations were also made explicit for an agent to ensure deterministic resolution of data paths and reduce reliance on implicit context.

The results across models, summarized in Figs.~\ref{fig:success_heatmap}–\ref{fig:error_distribution}, show consistent qualitative behavior with the baseline while highlighting quantitative differences in reliability, efficiency, and error patterns. For the \texttt{gemini-pro-2.5} model, the large number of repeated trials (219 total) provides a statistically robust characterization of performance across steps. For the other models - each tested with approximately ten trials per step - the smaller sample size limits statistical interpretation, and the results should therefore be regarded as qualitative indicators of behavior rather than precise performance estimates. Nonetheless, the observed cross-model consistency and similar failure patterns suggest that the workflow and evaluation metrics are sufficiently general to support larger-scale future benchmarks. This pilot-level comparison thus establishes both the feasibility and reproducibility of the agentic workflow across distinct LLM architectures.

Figure~\ref{fig:success_heatmap} summarizes the cross-model performance of the agentic workflow. The heatmap shows the success rate for each model–step pair, highlighting substantial variation in reliability across architectures. The dates appended to each model name denote the model release or update version used for testing, while the parameter in parentheses (e.g., ``17B'') indicates the model’s approximate parameter count in billions. Models in the \texttt{GPT-5} and \texttt{Gemini} families achieve the highest completion fractions across most steps, whereas smaller or open-weight models such as \texttt{gpt-oss-120b} exhibit lower but still non-negligible success on certain steps.
This demonstrates that the workflow can be executed end-to-end by multiple LLMs, though with markedly different robustness and learning stability.

\subsection{LLM-Driven Error Analysis}
Figure \ref{fig:error_distribution} shows the distribution of failure modes across all analysis stages, considering only trials that did not reach a successful completion. The categorization is based on the LLM-generated outputs themselves, capturing how each model typically fails. Error types include function-calling issues, missing or placeholder data, infrastructure problems (e.g., API or file-handling failures), and semantic errors - cases where the model misinterprets the prompt and produces runnable but incorrect code, as well as syntax mistakes.

During each run of the LLM agent workflow, two complementary log files are generated to assist in analysing and categorizing errors. An example of each log file can be found in Appendix \ref{appendix:log_files_and_error_analysis}.
\begin{itemize}
    \item \textbf{Comprehensive Log:} This file captures all LLM inputs and outputs across every call, as well as coder execution results.
    \item \textbf{Validation Log:} Here, the outputs produced by the LLM are compared against reference solutions, marked as failures if there are any structural mismatches (e.g., shape or dimension differences in numpy arrays) or content disparities. Outputs must match the solution exactly to pass. 
\end{itemize}

These log files serve as direct input for our LLM-based error classification pipeline. 

Error classification in our workflow is performed by a dedicated LLM (\texttt{gpt-oss-120b}), guided by a carefully engineered prompt to ensure standardized and unambiguous categorization. The prompt construction follows these principles:

\begin{itemize}
    \item \textbf{Agent Role:} The model is explicitly instructed to act as an expert on artificial intelligence workflow failures in high energy physics.
    \item \textbf{Workflow Summary:} The prompt summarizes the supervisor/coder workflow:
    \begin{itemize}
        \item The user provides an analysis prompt.
        \item A supervisor agent decomposes the task and instructs a coder agent.
        \item The coder agent generates code, which is then executed.
        \item The supervisor reviews execution outcomes and iterates with the coder as needed until completion.
    \end{itemize}
    \item \textbf{Error Categories:} The LLM receives the following list of possible error categories:
    \begin{itemize}
        \item \texttt{all data weights = 0}
        \item \texttt{dummy data created}
        \item \texttt{function-calling error}
        \item \texttt{incorrect branch name}
        \item \texttt{intermediate file not found}
        \item \texttt{semantic error}
        \item \texttt{other}
    \end{itemize}
    \item \textbf{Task Definition:} The prompt directs the model to select the \emph{single most appropriate} error category for each case, prioritizing root causes over superficial error messages. This means the categorization targets fundamental issues, such as logic mistakes, missing dependencies, data mismatches, or miscommunications. 
    \item \textbf{Strict Output Structure:} The model must return only the category name, wrapped in triple asterisks (e.g. \verb|***Category***|), with no additional explanation or content. This requirement allows for robust and automated parsing of the outputs.
    \item \textbf{Log File Inclusion:} Both the comprehensive and validation logs are provided to the LLM for context, ensuring informed and accurate error classification.
\end{itemize}

This prompt design ensures reproducible and interpretable error analysis, allowing us to systematically quantify and address failure modes within LLM-driven analysis workflows. 

Clear model-dependent patterns emerge. Models with higher success rates, such as the \texttt{GPT-5} and \texttt{Gemini 2.5} families, show fewer logic and syntax issues, while smaller or open-weight models exhibit more semantic and data-handling failures. These differences reveal characteristic failure signatures that complement overall completion rates. 

Taken together, Figure~\ref{fig:model_comparison} highlights two main aspects of LLM-agent behavior: overall task reliability and the characteristic error modes behind failed executions. These results show that models differ not only in success rate but also in the nature of their failures, providing insight into their robustness and limitations in realistic HEP workflows.
\begin{figure*}[htbp]
  \centering
  \includegraphics[width=0.95\textwidth]{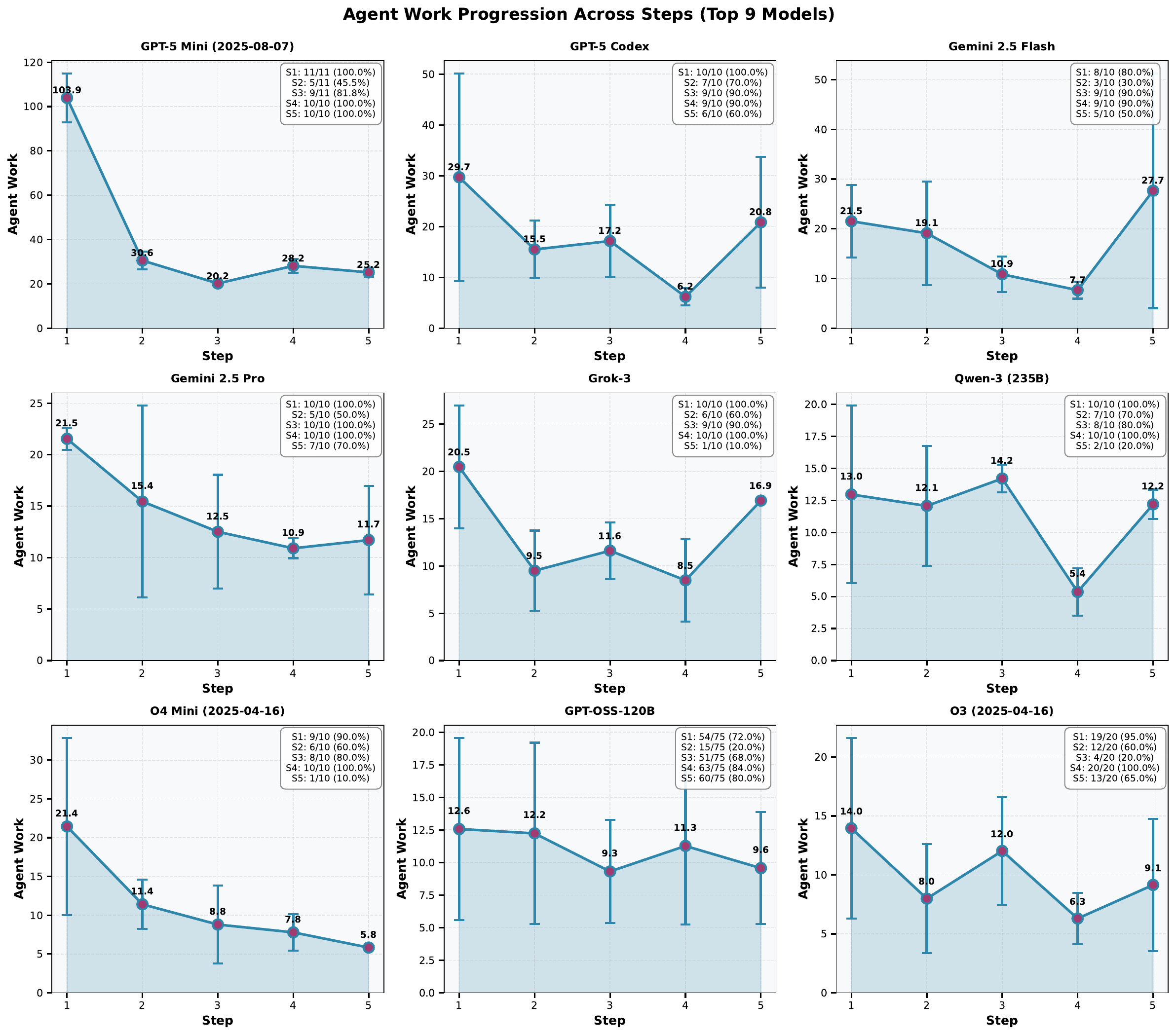}
  \caption{
    \textbf{Agent work progression across workflow steps.}
    Average agent work (with standard-deviation error bars) shown for the nine
    models that successfully completed all stages of the workflow.
  }
  \label{fig:agent_work}
\end{figure*}

\begin{figure*}[htbp]
  \centering
  \includegraphics[width=0.95\textwidth]{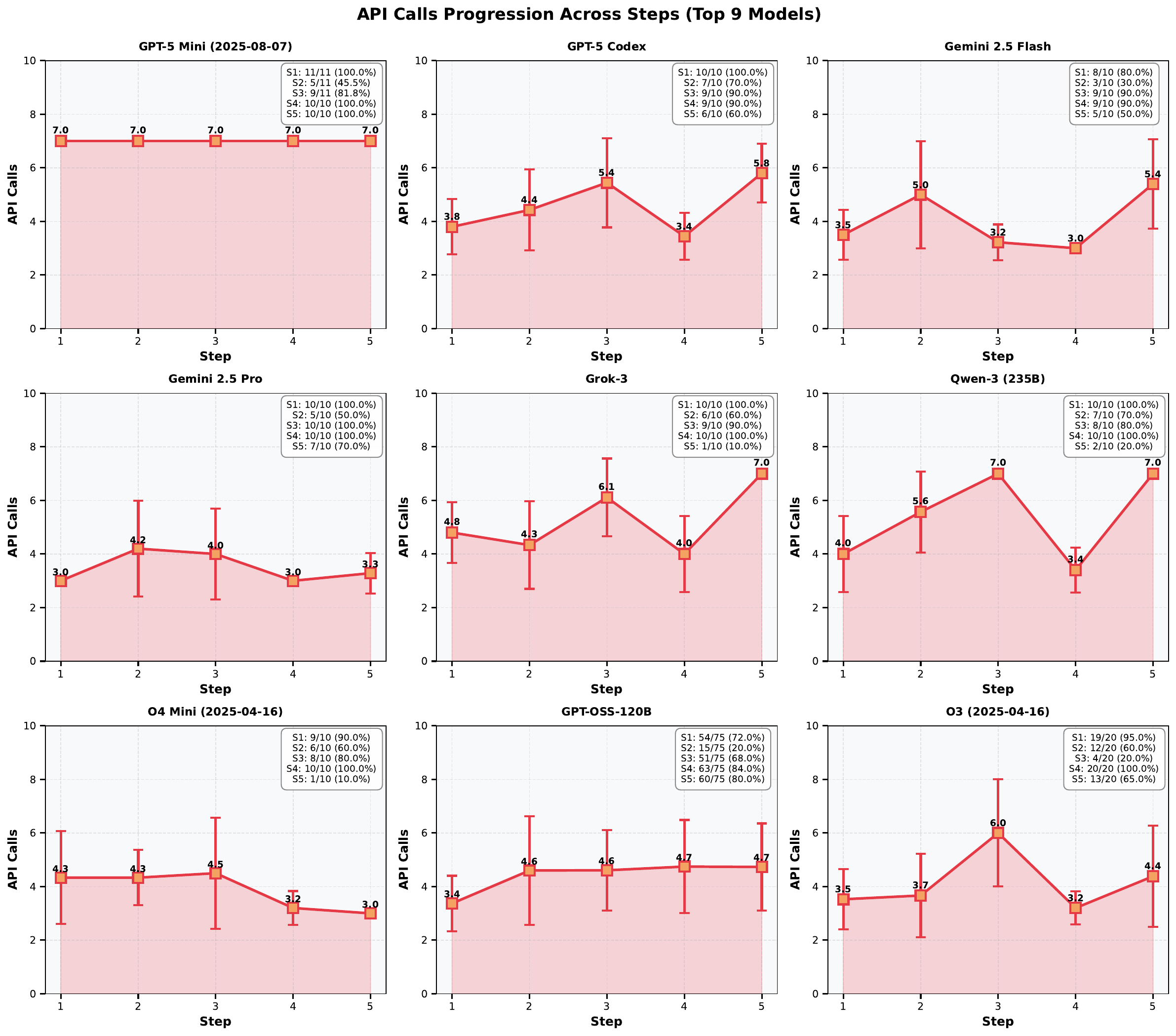}
  \caption{
    \textbf{API-call progression across workflow steps.}
    Average API-call counts (with standard-deviation error bars) for the nine
    models that completed all workflow stages.
  }
  \label{fig:api_calls_progression}
\end{figure*}

\subsection{Various Measures}
\begin{figure*}[htbp]
  \centering
  \includegraphics[width=0.95\textwidth]{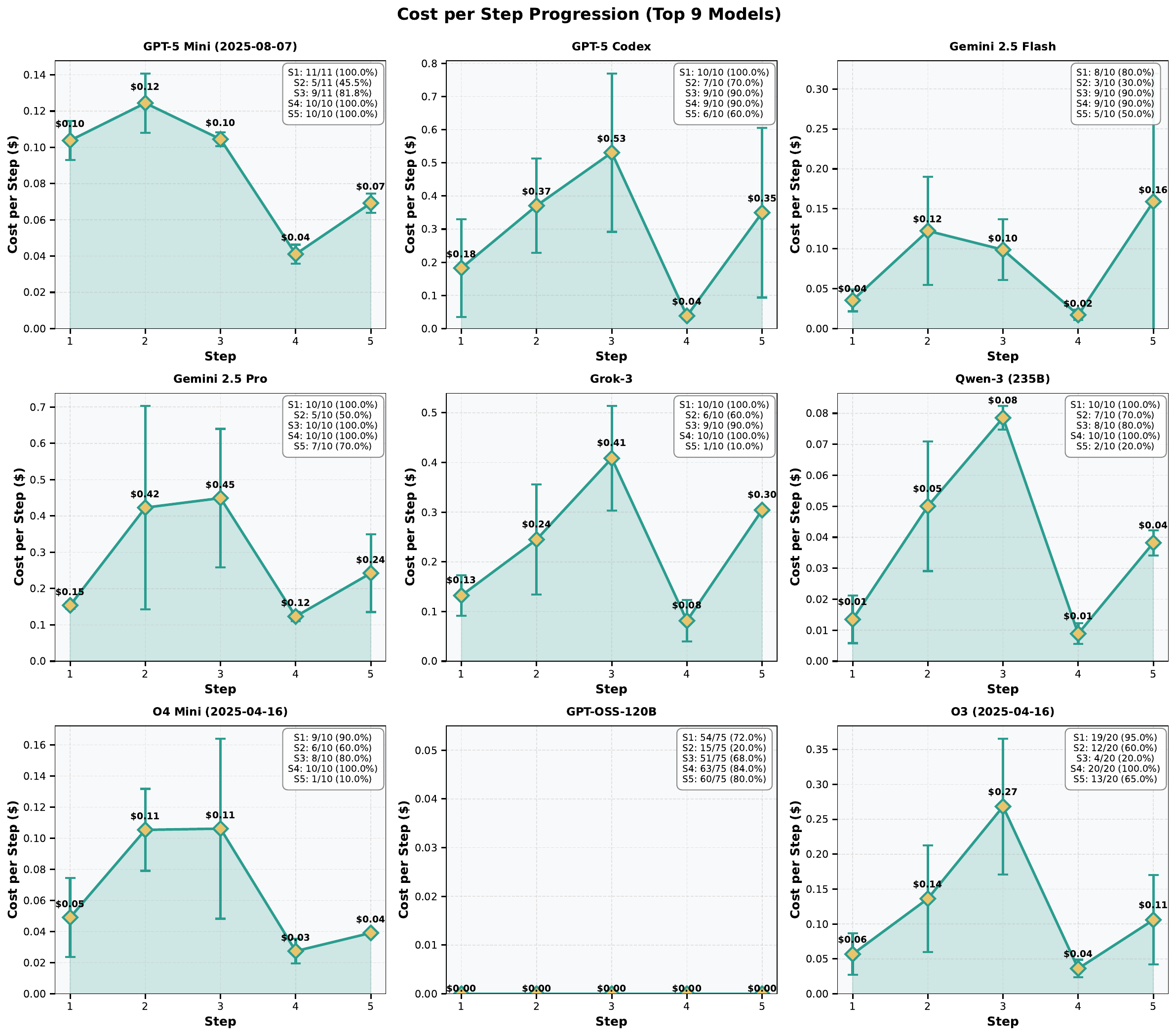}
  \caption{
    \textbf{Cost per analysis step across models.}
    The estimated dollar cost per step is computed from the observed token usage 
    and the publicly listed pricing for each model configuration, shown in Appedix~\ref{appendix:model_api_cost}. This metric 
    captures the effective economic cost of completing the workflow. 
    Higher-capacity models tend to combine lower agent work with higher dollar 
    cost, whereas smaller or open-weight models offer lower absolute cost but 
    substantially reduced reliability.
  }
  \label{fig:cost_per_step}
\end{figure*}
Figure~\ref{fig:agent_work}, Figure~\ref{fig:api_calls_progression}, and Figure~\ref{fig:cost_per_step} summarize results for a subset of the models tested. Although 20 models were evaluated in total (see Figure~\ref{fig:success_heatmap}), only nine achieved at least one successful completion in \emph{every} step of the workflow. We restrict the quantitative comparisons in these figures to this subset to ensure a fair and well-defined evaluation: metrics such as agent work, API-call behavior, and effective cost are meaningful only when a model is capable of completing the entire pipeline. Including models that never reached certain stages would bias these measurements and obscure true performance differences. The nine consistently successful models therefore constitute the evaluation set for the comparative analysis.

Figure~\ref{fig:agent_work} shows the agent work metric across workflow stages for the nine models that successfully completed all steps of the pipeline. Agent work is defined as
\begin{equation*}
    \text{Agent Work}\equiv\frac{\text{Total Tokens}-\text{User Tokens}}{\text{User Tokens}},
\end{equation*}
and measures the amount of autonomous reasoning, internal iteration, and self-correction performed beyond the initial user instruction.

The standard-deviation error bars quantify the stability of the model’s reasoning behavior across repeated trials. Models with small uncertainties exhibit highly consistent internal prompting patterns, while larger error bars indicate more volatile behavior—often corresponding to inconsistent self-repair strategies or sensitivity to intermediate outputs. These collective trends provide insight not only into how much iteration a model performs, but how reliably it executes those iterative processes.

The agent-work curves reveal substantial variation across both models and workflow stages. Several models show large bursts of internal prompting during the data-preparation step - consistent with the high failure rate observed for this stage - before settling into more stable behavior in later steps where the task structure becomes clearer. Other models display nearly uniform agent-work profiles, suggesting steadier, more systematic reasoning strategies across the entire workflow. Importantly, high success rates in Figure~\ref{fig:success_heatmap} do not necessarily correspond to low agent work. Some models achieve reliability through strong single-pass reasoning with minimal iteration, while others succeed only after extensive internal repair. This diversity highlights that low agent work is not inherently synonymous with robustness; rather, it reflects a particular style of problem-solving that may or may not be advantageous depending on task structure.


Figure~\ref{fig:api_calls_progression} reports the number of API calls issued by the agent at each workflow step for the nine models that successfully completed the entire pipeline. In our setup, an agent may attempt up to three corrective retries per step, corresponding to a maximum of seven API calls: one initial supervisor call followed by as many as three supervisor–coder interaction cycles. Whereas the agent-work metric captures internal reasoning effort, API-call counts measure how frequently the agent must rely on external interactions with the model to make progress.

The error bars quantify run-to-run variability and reveal differences in the stability of each model’s retry behavior. Some models maintain nearly constant API-call counts across all steps, indicating predictable correction strategies and consistent handling of intermediate failures. By contrast, several models exhibit higher API-call counts during the \textbf{categorization} stage, along with increased variability, suggesting that models find the optimization and decision-boundary logic in this stage particularly challenging. In comparison, data preparation generally requires fewer external interactions, reflecting its more structured and procedural character.

Overall, the API-call patterns highlight substantial diversity in how models manage uncertainty, recover from errors, and decide when external intervention is necessary. Some rely on frequent, lightweight retries, while others issue fewer but more targeted calls. These differences in external iteration behavior have direct implications for computational cost and operational efficiency, motivating the cost-per-analysis-step comparison presented next.

Figure~\ref{fig:cost_per_step} presents the estimated monetary cost per workflow step, computed by converting the observed token usage into dollars using publicly listed pricing for each model (Appendix \ref{appendix:model_api_cost}). Because each workflow stage involves different degrees of internal reasoning and external interaction, the resulting cost patterns provide insight into the computational demands of each step. Across nearly all models, the \textbf{preprocessing} stage is the most expensive. This reflects the substantial internal prompting, early error handling, and multi-step reasoning required for operations such as ROOT file inspection, ntuple construction, and initial event filtering. In contrast, the \textbf{S–B separation} step is consistently inexpensive, as it largely reduces to running a well-defined model-evaluation function with minimal iterative refinement. The \textbf{categorization} stage shows greater model-to-model variability: some models complete it with minimal additional iteration, while others incur noticeably higher cost due to repeated boundary adjustments and optimization retries.

The differences across models illustrate a clear capability–cost trade-off. Higher-capacity proprietary models often achieve lower agent work and more stable retry behavior, but these advantages come with higher per-token pricing, resulting in significantly elevated per-step cost. Smaller or open-weight models generally appear cheaper, but their reduced robustness can lead to inconsistent iteration patterns and occasional spikes in token usage. These trends emphasize that a model’s nominal price is only part of the story; the true operational cost depends on how efficiently the model navigates the reasoning and correction cycles required by each workflow stage.

A special case in Figure~\ref{fig:cost_per_step} is \texttt{gpt-oss-120b}, which appears with an effective cost close to zero. This stems from its execution on LBNL’s CBorg infrastructure~\citep{cborg}, where inference cost is extremely low (essentially zero). This highlights the appealing economic profile of open-weight models when deployed on institutional compute resources, though such low marginal cost does not necessarily imply reliability or performance comparable to commercial systems. Overall, the cost patterns show that the effective economic footprint of an agent depends jointly on model pricing and the internal and external iteration dynamics needed to successfully complete each stage of the scientific workflow.

\section{Limitations}
This study shows that LLMs can support HEP data analysis workflows by interpreting natural language, generating executable code, and applying basic self-correction. While multi-step task planning is beyond the current scope, the \texttt{Snakemake} integration provides a natural path toward rule-based agent planning. Future work will pursue this direction and further strengthen the framework through improvements in prompting, agent design, domain adaptation, and retrieval-augmented generation.

\section{Conclusion}
We demonstrated the feasibility of employing LLM agents to automate components of high-energy physics (HEP) data analysis within a reproducible workflow. The proposed hybrid framework combines LLM-based task execution with deterministic workflow management, enabling near end-to-end analyses with limited human oversight. While the \texttt{gemini-pro-2.5} model served as a statistically stable reference, the broader cross-model evaluation shows that several contemporary LLMs can perform complex HEP tasks with distinct levels of reliability and robustness. Beyond this proof of concept, the framework provides a foundation for systematically assessing and improving LLM-agent performance in domain-specific scientific workflows.

\section*{Acknowledgement}
This work was supported by the U.S. Department of Energy (DOE), Office of Science (SC), under Award No. DE-SC0023718.  Gendreau-Distler’s contribution was partially supported by the U.S. National Science Foundation under Award No. 2046280. Wang's contribution was partially supported by the DOE SC under the Contract No. DE-AC02-05CH11231. This research used the CBorg AI platform and resources provided by the IT Division at the Lawrence Berkeley National Laboratory (Supported by the Director, Office of Science, Office of Basic Energy Sciences, of the U.S.\ Department of Energy under Contract No.\ DE-AC02-05CH11231). The authors also thank Andrew Schmeder of LBNL’s IT Division for assistance with CBorg. This work additionally used resources of the National Energy Research Scientific Computing Center (NERSC), a DOE Office of Science User Facility operated under Contract No.\ DE-AC02-05CH11231. 

\newpage
\bibliographystyle{unsrtnat}
\bibliography{ref.bib}

\newpage
\appendix

\section{Technical Appendices and Supplementary Material}\label{appendix:physics_details}

\subsection{Features used for TabPFN classifier}
The language model is asked to save an extensive list of particle observables that would be useful for a more sophisticated physics analysis: the transverse momentum ($p_T$), pseudorapidity ($\eta$), and azimuthal angle ($\phi$) of the two leading photons, of the two leading leptons, and of the six leading jets; the transverse momentum and azimuthal angle of the missing transverse energy (MET); the Monte Carlo weight; the tight photon ID flags; the cross section; and the sum of Monte Carlo weights. However, in the interest of efficiency we use only the photon observables for the TabPFN classifier. The $p_T$ of photons are normalized by the invariant mass to prevent the classifier from learning to separate the signal and background events by the invariant mass. In addition, because only differences in azimuthal angle are physically meaningful, we replace the two diphoton $\phi$ coordinates with their difference $\phi_1 - \phi_2$.

\subsection{Event selection, training sample definition, and background estimation}

\paragraph{Preselection}  
To roughly follow the expected topology of a Higgs boson decaying into two photons, photon candidates are taken from well-calibrated regions of the detector ($|\eta| < 2.37$), while the transition zone between the barrel and endcap ($1.37 < |\eta| < 1.52$), where measurements are less reliable, is excluded. Each photon is required to have transverse energy above 25 GeV, ensuring that selected photons are energetic enough to be measured with good resolution and to suppress backgrounds from softer processes. Events are kept if they contain two photons that meet identification and isolation criteria, which help distinguish genuine photons from misidentified particles or nearby activity. To reflect the expected kinematics of Higgs decays, the leading photon must contribute more than 35\% of the diphoton system’s momentum ($p_T/m_{\gamma\gamma} > 0.35$), and the sub-leading photon more than 25\%. Finally, events are restricted to a diphoton invariant mass window of $105 < m_{\gamma\gamma} < 160$ GeV, a range chosen to capture the Higgs boson signal region while suppressing background contributions at much lower or higher masses.

\paragraph{Signal and background classifier training sample}  
Following the strategy of Ref.~\cite{ATLAS:2018mme}, events in which at least one photon does not meet the tight identification requirement are used to model the background in the signal region. These so-called NTI events are taken from the diphoton invariant mass sidebands, 105–120 GeV and 130-160 GeV, ranges chosen to minimize potential Higgs boson contributions. Because NTI events exhibit kinematic properties similar to those of tightly identified and isolated photons, they serve as a reliable proxy for background processes. For the signal sample, we use simulated Higgs boson events in which both photons satisfy the tight identification and isolation requirements. To ensure that the training is not biased toward either sample, signal and background events are rescaled to have equal weights.

\paragraph{Background estimate for sensitivity evaluation}  

To estimate the number of background events in the signal region (where both photons satisfy the tight isolation requirement), we rely on NTI events as a control sample. The signal is expected to appear as a narrow peak in the diphoton invariant mass distribution, so the significance is evaluated within a small mass window around the Higgs boson, $125 \pm 2$ GeV. The expected background in this window is inferred from the NTI sidebands using
\[
\text{Expected background} = SF1 \times SF2 \times NTI_{\text{yield}} .
\]  

In this expression, $NTI_{\text{yield}}$ denotes the number of NTI data events observed in the sideband regions. The scale factor $SF1$ accounts for the difference in yields between tight-isolated (TI) and NTI photons in the sidebands, while $SF2$ adjusts for the relative abundance of NTI events in the signal window compared with the sidebands. In this way, the measured background in the sidebands is extrapolated into the Higgs signal region.

\subsection{Procedure of event categorization}
The goal of the event categorization is to select boundaries among $1000$ evenly spaced values of the signal-background separation score in a way that maximizes the statistical significance of the Higgs-to-diphoton signal process. We adopt a number counting significance formula given by
\[{\textstyle Z_i = \sqrt{2\left[(s_i + b_i)\ln{\left(1+\frac{s_i}{b_i}\right)}-s_i\right]}, \quad Z_{\text{tot}} = \sqrt{\sum_i Z_i^2}}\]

where summation runs over all categories and \(s_i/b_i\) denote numbers of signal/background events in \(i^{\text{th}}\) category. To maximize this significance, the first step is to scan over all possible locations for the boundary, compute the significance with each candidate boundary, and ultimately keep the boundary that yields the highest statistical significance. This procedure is then repeated until the most recently added boundary improves the overall significance by less than $5\%$.

\begin{figure}[h]
\centering
\begin{minipage}{0.48\textwidth}
    \includegraphics[height=0.8\textwidth]{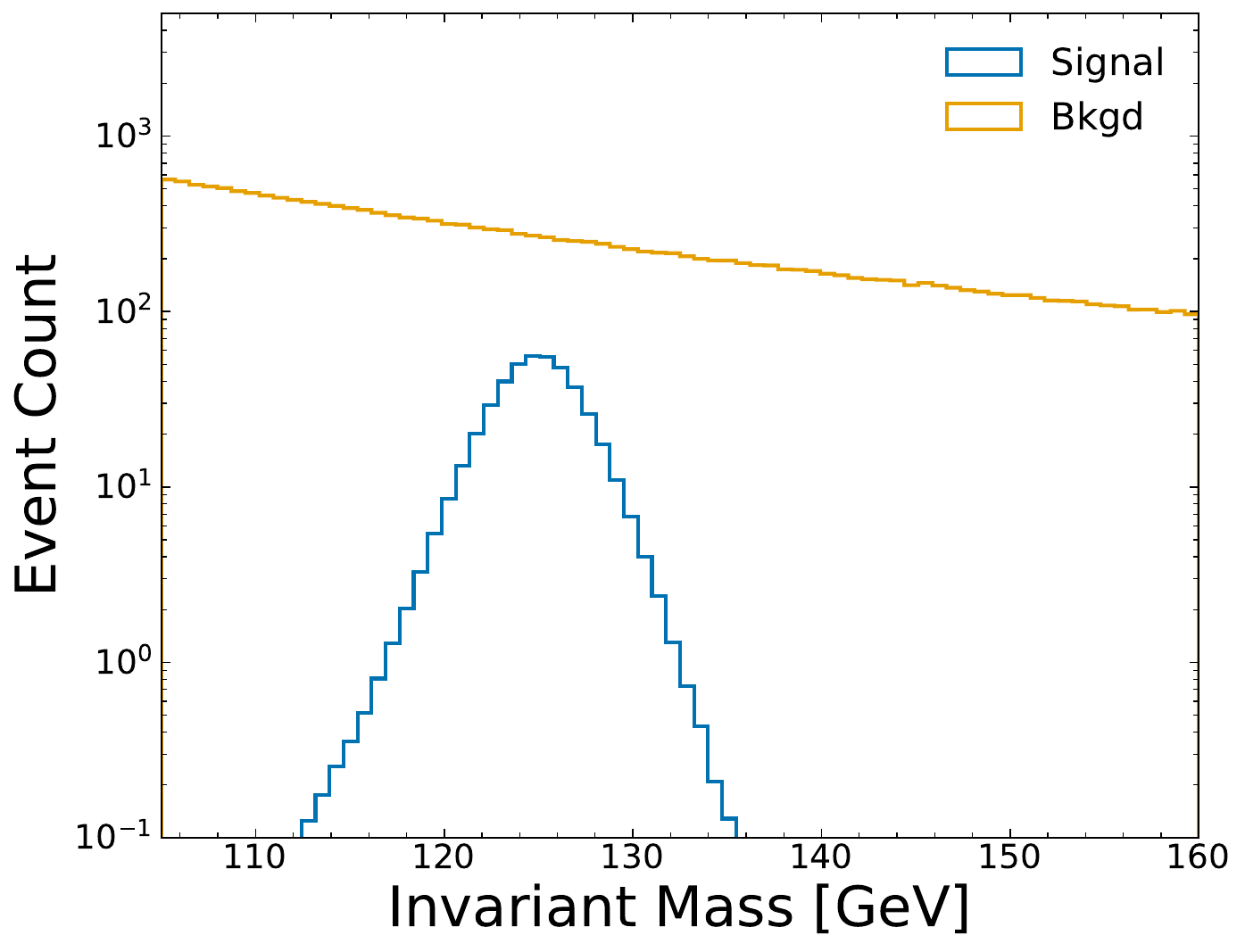}
\end{minipage}
\hfill
\begin{minipage}{0.48\textwidth}
    \includegraphics[height=0.8\textwidth]{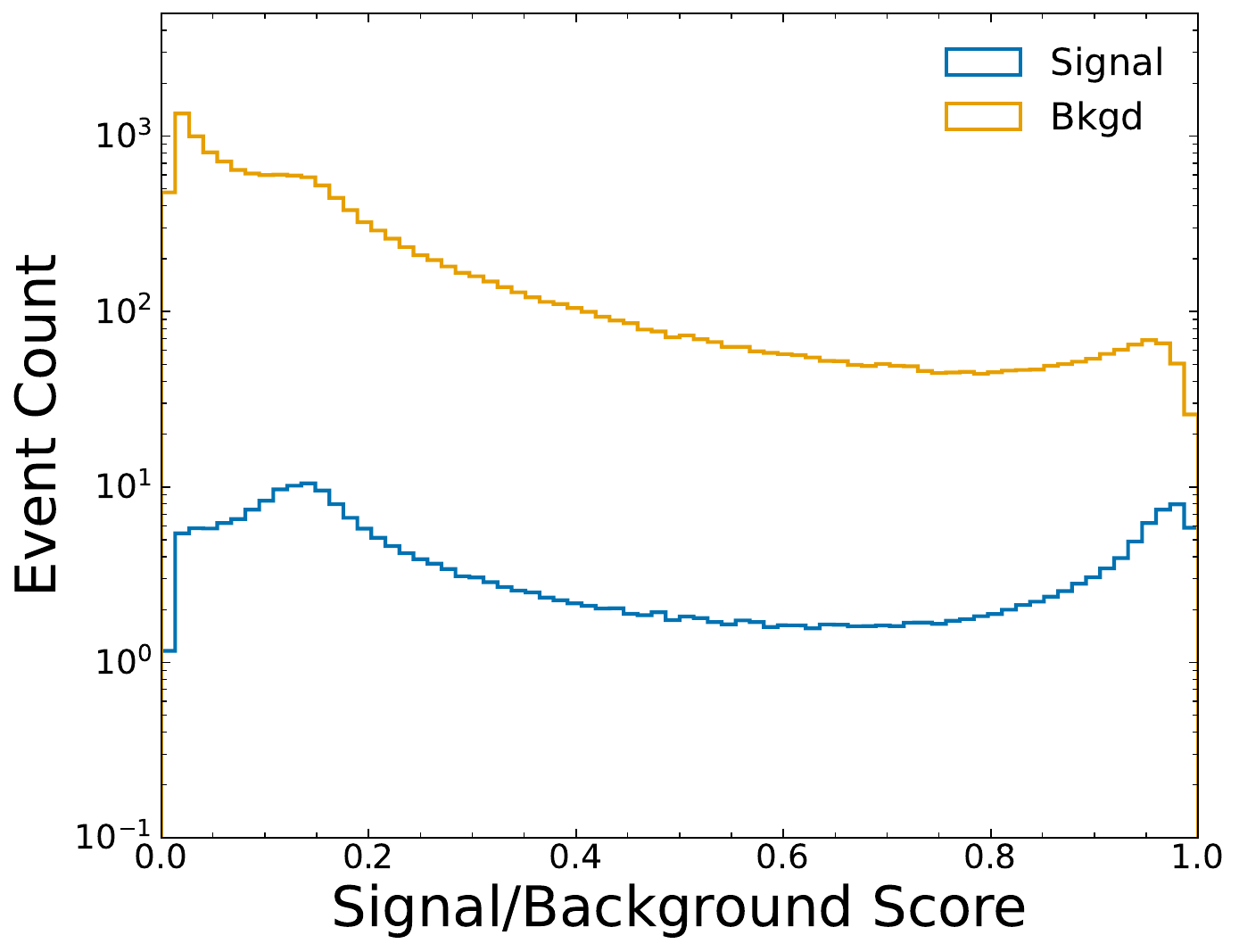}
\end{minipage}
\caption{Left: Invariant mass distribution of signal and background processes after standard analysis selections are applied. Right: Signal-background separation scores generated by TabPFN. Note that the normalization is different in the left and right plots because we apply cuts on the invariant mass before computing the signal-background separation scores.}
\label{fig:myy_scores}
\end{figure}

\section{Log Files and Error Analysis}\label{appendix:log_files_and_error_analysis}

\subsection{Comprehensive Log Example}

\begin{lstlisting}
==============================================
 COMPREHENSIVE SUPERVISOR-CODER LOG
==============================================

 Task: categorization
 Supervisor: anthropic/claude-sonnet:latest
 Coder: anthropic/claude-sonnet:latest
 Start Time: 2025-10-09 17:49:25.222518
 Working Directory: 
 Output Directory: 

 ORIGINAL USER PROMPT
--------------------------------------------------
Your task is to produce a set of boundaries that will categorize the provided samples in a way that maximizes the statistical significance.
The relevant samples are:
- Signal: 'solution/arrays/signal.npy'
- Background: 'solution/arrays/bkgd.npy'
- Signal scores 'solution/arrays/signal_scores.npy'
- Background scores: 'solution/arrays/bkgd_scores.npy'

Write a python script to produce the categorization using the following tools (headers provided below).
YOU MUST INCLUDE `from utils import *` in the script; do not attempt to write these functions yourself.

def load_datasets(signal, bkgd, signal_scores, background_scores):
    Return weighted and unweighted signal and background samples, signal_df and bkgd_df, as ROOT data frames.
    You must load the input arguments as np arrays before passing to the function.
    Example usage: signal_df, bkgd_df = load_datasets(signal, bkgd, signal_scores, bkgd_scores)

def get_significance(signal_df, bkgd_df, boundaries):
    Return significance under current categorization.
    Example usage: Z = get_significance(signal_df, bkgd_df, boundaries)

def place_boundary(signal_df, bkgd_df, boundaries, num_bins, min_events):
    Return optimal location to place next boundary based on current boundaries, and resulting significance.
    Example usage: new_boundary, new_Z = place_boundary(signal_df, bkgd_df, boundaries, num_bins, min_events)

Use the load_datasets(signal, bkgd, signal_scores, bkgd_scores) tool to get the signal and background histograms.
Keep track of the current categorization with an array containing the locations of the current boundaries (so start out with boundary_arr=[0,1]). num_bins should be set to 1000. Each time you want to place a boundary, use place_boundary to get the location of the new boundary and the resulting significance. Repeat until the significance improved by less than 5 percent as a result of adding the most recent boundary (that is, (new_significance - old_significance) / old_significance < 0.05). However, keep this last boundary computed for which the improvement in significances is less than 5%.

Save the boundary array to 'arrays/boundaries.npy' and the significance array (i.e., significance after adding each boundary) to 'arrays/significances.npy'.

 PROCESS START
--------------------------------------------------



=========  INITIALIZATION =========

[2025-10-09 17:49:25] INFO: Starting supervisor-coder interaction


=========  LOOP START =========

[2025-10-09 17:49:25] INFO: === MAIN LOOP STARTED ===


=========  SUPERVISOR CALL =========

[2025-10-09 17:49:25] INFO: Calling supervisor anthropic/claude-sonnet:latest with initial prompt


========= TOKEN STATS =========

[2025-10-09 17:49:40] INFO: Initial supervisor prompt tokens: 1418


========= SUPERVISOR RESPONSE =========

[2025-10-09 17:49:40] INFO: Supervisor response received (1418 input tokens, 925 output tokens)
## Worker Prompt

Write a Python script that optimizes categorization boundaries to maximize statistical significance for signal vs background classification. The script should:

1. Import necessary libraries including `numpy` and `from utils import *`

2. Load the input data:
   - Signal scores from: solution/arrays/signal_scores.npy`
   - Background scores from: solution/arrays/bkgd_scores.npy`
   - Signal samples from: solution/arrays/signal.npy`
   - Background samples from: solution/arrays/bkgd.npy`

3. Use `load_datasets(signal, bkgd, signal_scores, bkgd_scores)` to create signal_df and bkgd_df ROOT dataframes

4. Initialize boundaries array as `[0, 1]` and an empty list to track significances

5. Set `num_bins = 1000` and `min_events` to an appropriate value (you can use a default like 10)

6. Iteratively add boundaries:
   - Use `place_boundary(signal_df, bkgd_df, boundaries, num_bins, min_events)` to find the next optimal boundary
   - Add the new boundary to the boundaries array (keep sorted)
   - Record the new significance
   - Calculate the relative improvement: `(new_significance - old_significance) / old_significance`
   - Continue until the improvement is less than 0.05 (5%)
   - IMPORTANT: Keep the last boundary even if it does not meet the 5% improvement threshold

7. Save outputs:
   - Boundaries array to: arrays/boundaries.npy`
   - Significances array to: arrays/significances.npy`

Return complete, executable Python code.

## Call record:

Requested worker to create a Python script that implements an iterative boundary optimization algorithm for signal/background categorization. The script will:
- Load signal and background samples and their scores from the specified input files
- Use the provided utility functions (load_datasets, place_boundary, get_significance) to iteratively find optimal category boundaries
- Start with boundaries [0, 1] and add new boundaries one at a time
- Continue adding boundaries until the significance improvement drops below 5%
- Save the final boundaries and significance progression to the specified output files

Input files:
solution/arrays/bkgd_scores.npy
solution/arrays/signal_scores.npy
Output files:
arrays/boundaries.npy
arrays/significances.npy


=========================  CODER CALL #1 =========================

[2025-10-09 17:49:40] INFO: Calling coder anthropic/claude-sonnet:latest (attempt 1/3)


====================== CODER RESPONSE #1 ======================

[2025-10-09 17:49:51] INFO: Coder response received (810 input tokens, 985 output tokens)
Generated Code:
```python
import numpy as np
from utils import *

# Load the input data
signal_scores = np.load('solution/arrays/signal_scores.npy')
bkgd_scores = np.load('solution/arrays/bkgd_scores.npy')
signal = np.load('solution/arrays/signal.npy')
bkgd = np.load('solution/arrays/bkgd.npy')

# Create signal_df and bkgd_df ROOT dataframes
signal_df, bkgd_df = load_datasets(signal, bkgd, signal_scores, bkgd_scores)

# Initialize boundaries and significances
boundaries = [0, 1]
significances = []

# Set parameters
num_bins = 1000
min_events = 10

# Iteratively add boundaries
while True:
    # Find the next optimal boundary
    new_boundary, new_significance = place_boundary(signal_df, bkgd_df, boundaries, num_bins, min_events)
    
    # Add the new boundary to the boundaries array (keep sorted)
    boundaries.append(new_boundary)
    boundaries.sort()
    
    # Record the new significance
    significances.append(new_significance)
    
    # Calculate the relative improvement
    if len(significances) > 1:
        old_significance = significances[-2]
        relative_improvement = (new_significance - old_significance) / old_significance
        print(f`Iteration {len(significances)}: Boundary = {new_boundary:.4f}, Significance = {new_significance:.4f}, Improvement = {relative_improvement:.4f}`)
        
        # Check if improvement is less than 5%
        if relative_improvement < 0.05:
            print(f`Stopping: Improvement ({relative_improvement:.4f}) is less than 0.05`)
            break
    else:
        print(f`Iteration {len(significances)}: Boundary = {new_boundary:.4f}, Significance = {new_significance:.4f}`)

# Convert significances to numpy array
significances = np.array(significances)

# Save outputs
np.save('arrays/boundaries.npy', boundaries)
np.save('arrays/significances.npy', significances)

print(f`\nFinal boundaries: {boundaries}`)
print(f`Final significances: {significances}`)
print(f`Total number of boundaries: {len(boundaries)}`)
```


=================== CODE EXECUTION #1 ====================

[2025-10-09 17:49:51] INFO: Executing generated code.
[2025-10-09 17:49:51] DEBUG: Command: python generated_code/categorization.py --input-files solution/arrays/bkgd_scores.npy solution/arrays/signal_scores.npy --output-files arrays/boundaries.npy arrays/significances.npy


===================== EXECUTION RESULT #1 =====================

[2025-10-09 17:51:02] INFO: Execution completed with exit code 0
[2025-10-09 17:51:02] INFO: Execution successful. Output:
Iteration 1: Boundary = 0.8640, Significance = 0.6860
Iteration 2: Boundary = 0.9780, Significance = 0.8987, Improvement = 0.3101
Iteration 3: Boundary = 0.1170, Significance = 0.9359, Improvement = 0.0414
Stopping: Improvement (0.0414) is less than 0.05

Final boundaries: [0, 0.117, 0.864, 0.978, 1]
Final significances: [0.68595282 0.89867695 0.93591648]
Total number of boundaries: 5



=================  SUPERVISOR FEEDBACK CALL #1 =================

[2025-10-09 17:51:02] INFO: Calling supervisor anthropic/claude-sonnet:latest for feedback (iteration 1)


================ SUPERVISOR FEEDBACK RESPONSE #1 ================

[2025-10-09 17:51:04] INFO: Supervisor feedback received (2832 input tokens, 10 output tokens)
Supervisor Feedback:
Supervisor is satisfied with current results


=========================  SUCCESS =========================

[2025-10-09 17:51:04] INFO: SUCCESS: Supervisor satisfied after 1 iterations!


===================  FINAL STATISTICS ===================

[2025-10-09 17:51:04] INFO: Total API calls made: 3
[2025-10-09 17:51:04] INFO: Final token counts: 5060 input, 1920 output


================  PROCESS COMPLETED ================

[2025-10-09 17:51:04] INFO: Interaction completed in 0:01:38.951148


=========================  END OF LOG =========================

[2025-10-09 17:51:04] INFO: === END OF COMPREHENSIVE LOG ===

==========================================
 FINAL SUMMARY
==========================================
Task: categorization
Supervisor: anthropic/claude-sonnet:latest
Coder: anthropic/claude-sonnet:latest
Total Iterations: 1
Start Time: 2025-10-09 17:49:25.222518
End Time: 2025-10-09 17:51:04.173666
Duration: 0:01:38.951148
Total API Calls: 3
Total Input Tokens: 5060
Total Output Tokens: 1920
User Prompt Tokens: 1418
Supervisor to Coder Tokens: 810
Coder Output Tokens: 985
Feedback to Supervisor Tokens: 2832
Final Status: SUCCESS
==========================================
\end{lstlisting}

\subsection{Validation Log Example}

\begin{lstlisting}
=== Step 5 Validation (categorization) ===
Validating boundaries.npy...
  [X] Shape mismatch: (5,) vs (4,)
Validating significances.npy...
  [X] significances.npy differs: NaN mask equal=True, finite mismatches=2/3
     diff stats: max=0.0390664, mean=0.0254353

Running detailed mismatch analysis...

Detailed analysis for significances.npy (using 4 significant digit tolerance):
  Generated shape: (3,), Reference shape: (3,)
  Tolerance: Adaptive based on 4 significant digits per value
  [X] 2 elements differ (out of 3 total)
  Sample numeric mismatches (first 10 indices):
    Index 1: gen=0.898676950456262, ref=0.8596105777616457, diff=0.039066372694616214
    Index 2: gen=0.935916477251495, ref=0.898676950456262, diff=0.03723952679523301
  1D array - no column-by-column analysis needed

=== VALIDATION SUMMARY ===
Step: 5 (categorization)
Files validated:
  - boundaries.npy - Category boundary thresholds
  - significances.npy - Statistical significance values

Step 5 (categorization): FAIL
[X] Validation failed

Result: 0/1 step passed
\end{lstlisting}

\subsection{LLM-Error Analysis Example}
From the comprehensive log, we observe that the Supervisor/Coder was tasked with determining categorization boundaries to split events based on a classifier score. However, validation revealed that the boundaries selected did not exactly match those specified in the provided solution, leading to a failed trial. According to the LLM-based error analysis, this was classified as a \texttt{semantic error}, attributed to miscommunication between the User, Supervisor, and Coder. Specifically, the Supervisor misinterpreted the user's intent, and the Coder implemented instructions accordingly. 

A crucial difference emerged between the code generated by the Supervisor/Coder and the reference solution, namely, the solution initialized the \texttt{significances} array with the significance of the null selection (category boundaries at 0 and 1). In contrast, the Supervisor/Coder's approach did not include this initial step. This discrepancy led the LLM agent to add an extra boundary not present in the solution, ultimately producing five categorization boundaries instead of the four expected. As a result, the trial was marked as unsuccessful due to a semantic error.

\section{Model API Cost}\label{appendix:model_api_cost}
\begin{table}[htbp]
\centering
\begin{tabular}{|l|c|c|}
\hline
\textbf{Model} &
\shortstack{ \\ \textbf{Input Cost} \\ (\$/1M tokens)} &
\shortstack{\\ \textbf{Output Cost} \\ (\$/1M tokens)} \\
\hline
Claude 3.5 Haiku (2024-10-22)      & 0.80   & 15.00 \\
Claude Haiku 4.5 (2025-10-01)      & 1.00   & 5.00  \\
Claude Opus 4.1 (2025-08-05)       & 15.00  & 75.00 \\
Claude Sonnet 4.5 (2025-09-29)     & 3.00   & 15.00 \\
DeepSeek-R1                        & 0.28   & 0.42  \\
GPT-5 (2025-08-07)                 & 1.25   & 10.00 \\
GPT-5 Codex                        & 1.25   & 10.00 \\
GPT-5 Mini (2025-08-07)            & 0.25   & 2.00  \\
GPT-OSS-120B                       & 0.00   & 0.00    \\
Gemini 2.0 Flash Lite              & 0.10   & 0.40  \\
Gemini 2.5 Flash                   & 0.30   & 2.50  \\
Gemini 2.5 Pro                     & 1.25   & 10.00 \\
Grok Code Fast 1                   & 0.20   & 1.50  \\
Grok Mini                          & 0.30   & 0.50  \\
Grok-3                             & 3.00   & 15.00 \\
Llama-4 Maverick (17B)             & 0.50   & 2.00    \\
Llama-4 Scout (17B)                & 0.30   & 1.00    \\
O3 (2025-04-16)                    & 2.00   & 8.00  \\
O3 Mini (2025-01-31)               & 1.10   & 4.40  \\
O4 Mini (2025-04-16)               & 1.10   & 4.00  \\
Qwen-3 (235B)                      & 0.50   & 2.00    \\
\hline
\end{tabular}
\caption{Input and output costs for each model (in US dollars per 1M tokens). All cost values reflect the most recent updates as of October 2025.}\label{tab:model_api_cost}
\end{table}

\end{document}